\documentclass{aa}  

\usepackage{graphicx}
\usepackage{txfonts}
\usepackage{lipsum}
\usepackage{subcaption}         % necessary for continued figures, example in section 3
\usepackage{lscape}             % to rotate a single page table, example in appendix.
\usepackage{placeins}           % useful with \FloatBarrier, to keep 
\usepackage[backref, pagebackref=false, hyperindex=true, breaklinks=true, colorlinks=true, urlcolor=magenta, linkcolor=magenta, citecolor=NavyBlue, citebordercolor={0 1 0}, pagecolor=red, bookmarks=true, filecolor=blue,bookmarksopen=true, plainpages=false, pdfpagemode=UseThumbs]{hyperref}
\usepackage[dvipsnames]{xcolor}
\usepackage{ulem}

\usepackage[]{ulem}  % to use sout command

\usepackage{hyperref}
\usepackage{subcaption}
\usepackage{caption}

\begin{document}

   \title{Towards a global model for planet formation in layered MHD wind-driven discs}

   \subtitle{A population synthesis approach to investigate the impact of low viscosity and accretion layer thickness}

%%%%%%%%%%%%%%%%%%%%%%%%%%%%%%%%%%%%%%%%
% Please do not include ORCIDs next to author names.
% Only ORCIDs authenticated by individual authors in EDP Sciences editorial system will be taken into account.
% ORCIDs included here will be removed.
%%%%%%%%%%%%%%%%%%%%%%%%%%%%%%%%%%%%%%%%

\author{Jesse Weder \inst{\ref{unibe}}
        \and
        Christoph Mordasini \inst{\ref{unibe},\ref{CSH}}
        }

\institute{
        Division of Space Research and Planetary Sciences, Physics Institute, University of Bern, Gesellschaftsstrasse 6, 3012 Bern, Switzerland\\ \email{jesse.weder@unibe.ch} \label{unibe}
        \and
        Center for Space and Habitability, University of Bern, Gesellschaftsstrasse 6, 3012 Bern, Switzerland\label{CSH}
        }
\date{Received 16 September 2025; accepted 19 December 2025}

% \abstract{}{}{}{}{}
% 5 {} token are mandatory
  \abstract
   {Planet formation is inherently linked to the evolution of the protoplanetary disc. Recent developments point towards discs evolving due to magnetised winds instead of turbulent viscosity. This has fundamental implications for planet formation.}
   {We investigate planet formation in the context of magnetohydrodynamic (MHD) wind-driven disc evolution under the assumption of accretion being driven in a laminar accretion layer at the disc surface above a disc midplane with low turbulent viscosity. We aim at testing the global consequences of recent findings from 2D and 3D hydrodynamical simulations regarding inefficient midplane heating and the existence of two sub-regimes of Type II migration, namely slow viscosity-dominated and fast wind-driven migration.}
   {To study the global, potentially observable imprints of the physical processes governing planet formation in layered MHD-wind-driven discs, we run single-embryo planetary population syntheses with varying initial disc conditions (i.e. disc mass, size and angular momentum transport) and embryo starting location. We test different parametrisations for the accretion layer thickness $\Sigma_\mathrm{active}$.}
   {The extent of Type II migration in layered discs depends sensitively on the considered accretion layer thickness. For thin ($\Sigma_\mathrm{active}\lesssim0.01\,\mathrm{g/cm^2}$ or fast ($\gtrsim12\,\%$ sonic velocity) accretion layers, giant planets migrate in the slow viscosity-dominated regime which strongly limits the extent of Type II migration. The fast wind-driven sub-regime nearly never not occurs. For thick ($\Sigma_\mathrm{active}\gtrsim1\,\mathrm{g/cm^2}$) or slow ($\lesssim3\,\%$ sonic velocity) accretion layers, fast-wind driven type II occurs in contrast frequently, leading to long-range inward migration that sets in once planets reach masses sufficient to block the accreting layer (typically several $100\,\mathrm{M_\oplus})$. Disk-limited gas accretion is also strongly affected by deep and early gap opening, limiting  maximum giant planet masses.}
   {The existence of two subtypes of Type II migration, low Type I - Type II transition masses and limited runaway gas accretion in layered MHD wind-driven disks strongly influence the final mass–distance diagram of planets. For thin layers, giant planets form nearly in situ once they have passed into type II migration, which happens already at a few Earth masses. This leads to a bifurcation of the formation tracks where low-mass planets (super-Earths and sub-Neptunes) form closer in while giant planets remain farther out in the disc $\gtrsim 1\,\mathrm{au}$. For thick layers, fast wind-driven migration leads in contrast to numerous migrated Hot Jupiters. Overall, we find that while the global properties of the emerging planet population are strongly modified relative to classical viscous discs, key properties of the observed population can be reproduced within this new paradigm.}

   \keywords{protoplanetary disks -- magnetohydrodynamics (MHD) -- planets and satellites: formation -- planet-disk interactions}

   \maketitle
\nolinenumbers

%===============
\section{Introduction}
%===============

%FFFFFFFFFFFFFF
\begin{figure*}[!ht]
    \centering
    % trim: left bottom right top
    \includegraphics[trim={20cm 17cm 0 10cm},clip,width=\textwidth]{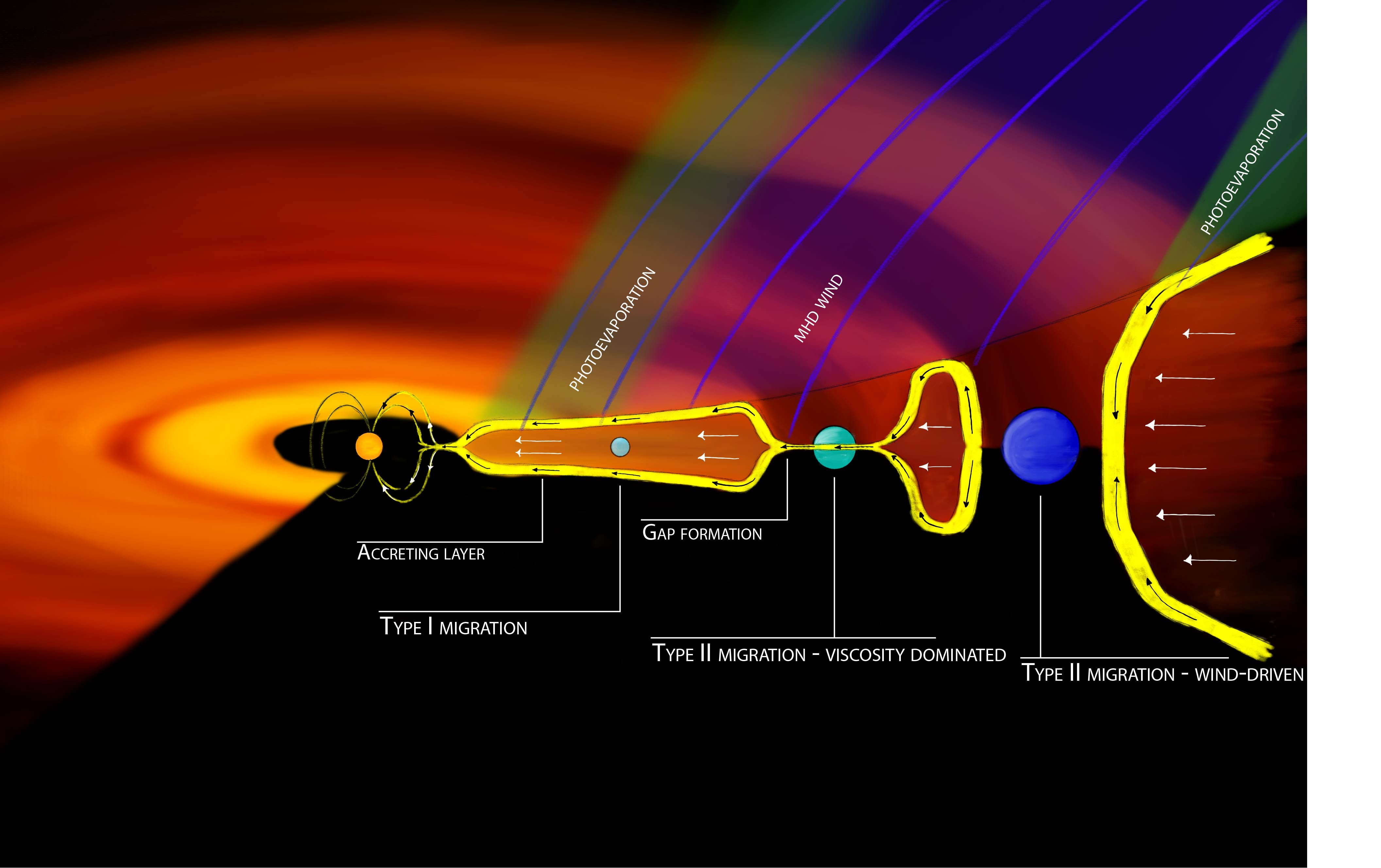}
    \caption{Sketch of the global framework explored in this work. Disc evolution through angular momentum extraction by an MHD wind with additional low-level internal and external photoevaporation with accretion occurring in layers at the disc surface. Low-mass planets remain embedded and undergo Type I migration. Growing planets eventually open a gap and transition to Type II migration. Initially, the fast accretion layer bypasses the planet, which migrates due to Lindblad torques from the gap (viscosity-dominated regime) while still accreting. At higher masses, the planet fully blocks the flow, entering wind-driven Type II migration, where it is pushed by the outer gap replenished by the accretion flow.}
    \label{fig:concept_overview}
\end{figure*}
%FFFFFFFFFFFFFF

As of today more than 6000 confirmed exoplanets have been discovered with many more candidates\footnote{As of August 2025, according to \href{https://exoplanet.eu}{https://exoplanet.eu}.} This population of extrasolar planets shows a wide diversity in planetary masses and orbital distances. The formation pathways of these different planets is subject to current studies, however the multitude of physical processes involved and the fact that they are interlinked complicates matter. These physical processes are often not fully understood, difficult or impossible to be observed. Using global models that include many of the physical processes that are believed to be of fundamental importance our current understanding of planet formation can be put to the test \cite[see review by][]{BurnMordasini2024}. The approach of so called planetary population syntheses was first introduced by \cite{IdaLin2004a} and refined in succeeding works \citep[e.g.][]{Alibert2004,Alibert2005,Mordasini2009a,Mordasini2009b, Alibert2013,Bitsch2015,Brugger2018,Emsenhuber_NGPPSI_2021,Emsenhuber_NGPPSII_2021,Alessi2018,Alessi2022,Speedie2022,Drazkowska2023,Kimura2022}, incorporating more recent developments in planet formation theory.

Planet formation is inherently close related to the evolution of the protoplanetary disc as the disc is delivering the material from which planets grow \citep{Miotello_PPVII_2023}. However, the exact processes driving the evolution of the disc remain unknown \citep{Manara_PPVII_2023}. In the last decade, the paradigm has shifted towards discs being driven by magnetohydrodynamic winds (MHD) that can efficiently remove angular momentum \citep{BlandfordPayne1982,BaiStone2013}. Pioneering work by \cite{Ogihara2015a,Ogihara2015b} showed that the inclusion of disc winds can help to suppress fast Type I inwards migration. \cite{Speedie2022} investigated the effect of a low viscosity disc with an outer turbulent region and found a bifurcation depending on the varying disc viscosity. \cite{Alessi2022} continued investigating planet formation under the combined effects of radially varying angular momentum transport by turbulent viscosity and disc winds using the planet population syntheses approach and found such a bifurcation to be persisting.

More recent 2D and 3D (magneto) hydrodynamic simulations including the effects of an MHD wind show the emergence of various new effects on planet formation via the disc temperature structure \cite[e.g.][]{Mori2019,Mori2021,Mori2025}, gap opening \cite[e.g.][]{Elbakyan2022,AoyamaBai2023,Wafflard-Fernandez2023,Hu2025}, migration \cite[e.g.][]{Kimmig2020,McNally2020,Lega2022,AoyamaBai2023,Wafflard-Fernandez2023,Wafflard-Fernandez2025,Hu2025,Wu2025} and gas accretion \citep[e.g.][]{Nelson2023,Hu2025}.

This work focuses on connecting some of these aspects in a global picture and to assess the potential impact of MHD wind-driven disc evolution on our current understanding of planet formation. We investigate the core accretion scenario within an MHD wind-driven disc where accretion is occurring in accretion layers at the disc surface. Planetary embryos inserted in the disc grow due to pebble accretion, remaining embedded and migrating in type I regime, not influenced by the accretion layer. As the planets continue to grow due to gas accretion, they eventually open a gap. However, the accretion flow is still able to flow through the gap and the planets migrate due to the Lindblad torque inside the gap. We call this the viscosity-dominated type II regime. If the planets continue to grow they eventually are able to block the accretion flow and are then pushed by the outer gap being replenished from the accretion flow. This picture of Type II migration follows largely what was found by \cite{Lega2022}.
Fig.~\ref{fig:concept_overview} gives a schematic view of the global picture investigated. In Sect.~\ref{sec:methods} we describe the details of the model and show exemplary cases. We then show and discuss the results of single-embryo planetary population syntheses (SEPPS) in Sect.~\ref{sec:results}.

%===============
\section{Methods} \label{sec:methods}
%===============
The model is based on the Bern Model of Planet Formation and Evolution \citep{Alibert2005,Mordasini2012,Emsenhuber_NGPPSI_2021} including recent findings regarding planet formation in MHD wind-driven discs. The disc evolution is introduced and discussed in Sect.~\ref{subsec:disc_model}. Planet-disc interaction such as migration and gap-opening are described in Sect.~\ref{subsec:migration_model} and the planetary growth model is introduced in Sect.~\ref{subsec:planet_growth}.

%---------------
\subsection{Disc evolution model} \label{subsec:disc_model}
%---------------
The evolution of the gas disc is calculated using the model in \cite{Weder2023}, which includes MHD wind-driven disc evolution based on \cite{Suzuki2016}.

%- initial gas profile
We start from an initial gas surface density profile
\begin{equation}
    \Sigma_\mathrm{g,init}(r)=\Sigma_\mathrm{g,0,5.2au}\left( \frac{r}{5.2\,\mathrm{au}} \right)^{-\beta} \exp \left[ \left( -\frac{r}{R_\mathrm{char}} \right)^{2-\beta} \right] \left( 1 - \sqrt{\frac{R_\mathrm{in}}{r}} \right)
\end{equation}
with power-law index $\beta$, $R_\mathrm{in}$ being the inner disc edge and characteristic radius $R_\mathrm{char}$. $\Sigma_\mathrm{g,0,5.2au}$ is the suitable value such that the disc mass $M_\mathrm{disc}$ corresponds to the assigned value.

%- gas disc evolution
The time evolution of the integrated gas surface density $\Sigma_\mathrm{g}$ is given by the advection-diffusion equation
\begin{equation} \label{eq:disc_evolution}
   \begin{split}
      \frac{\partial \Sigma_\mathrm{g}}{\partial t} =& \frac{1}{r}\frac{\partial}{\partial r}\left[ \frac{3}{r\Omega}\frac{\partial}{\partial r}(r^2\Sigma_\mathrm{g} \overline{\alpha_{r\phi}} c_\mathrm{s}^2) \right] + \frac{1}{r}\frac{\partial}{\partial r}\left[ \frac{2}{\Omega}r \overline{\alpha_{\phi z}}(\rho c_\mathrm{s}^2)_\mathrm{mid} \right] \\
      &- \dot{\Sigma}_\mathrm{MDW} - \dot{\Sigma}_\mathrm{PEW,int} - \dot{\Sigma}_\mathrm{PEW,ext},
  \end{split}
\end{equation}
with $r$ being the distance from the central star, $\Omega$ corresponding to the Keplerian frequency and $c_\mathrm{s}$ and $\rho$ being sound speed and density at the disc midplane. The sink terms $\dot{\Sigma}$ correspond to various outflows (i.e. magnetic disc wind (MDW), internal- and external photoevaporation (PEW)). The stress acting on the disc is parameterised through $\overline{\alpha_{r\phi}}$ and $\overline{\alpha_{\phi z}}$. The $r \phi$ component accounts for radial transport of angular momentum through turbulence \cite[i.e. magnetorotational instability (MRI)][]{BalbusHawley1991,BalbusHawley1998} and the $\phi z$ component accounts for vertical extraction of angular momentum through the MHD wind. For simplicity, we here assume radially and temporally constant $\overline{\alpha_{r \phi}}$ and $\overline{\alpha_{\phi z}}$. The mass-loss rate through the magnetised wind $\dot{\Sigma}_\mathrm{MDW}$ is constrained by the fraction of released gravitational energy going into launching the wind \citep[for details about the energetic constraint see][]{Suzuki2016}. Following the results of \cite{Weder2023} we focus on the weak wind approach, where a fraction of the released gravitational energy is going into launching the wind ($1-\epsilon_\mathrm{rad}=0.9$), following \cite{Mori2025}. The remaining 10\% contribute to disc heating or are radiated away.

The assumption of a spatially and temporally constant $\overline{\alpha_\mathrm{\phi z}}$ represents a gross simplification. The stress from the MHD wind is related to the disc magnetisation $\beta \equiv P_\mathrm{th}/P_\mathrm{mag}$ (i.e. the ratio of thermal- and magnetic pressure) and evolves with the gas density $\Sigma_\mathrm{g}$ and magnetic field strength $B$ \citep[e.g.][]{BaiStone2013,Lesur2021}. However, the evolution of the magnetic field remains poorly constrained and we lack an analytical formulation. The assumption of a temporally constant $\overline{\alpha_\mathrm{\phi z}}$ made here, corresponds to a constant magnetisation $\beta$ and hence a magnetic field that dissipates at the same rate as the disc. The model used here is similar to the hybrid wind solution with $\Psi>1$ and constant $\alpha_\mathrm{DW}$ in \cite{Tabone2022} (see their Fig.~6). The evolution of the magnetic torque has a strong impact on the evolution of the surface density in the inner disc, and can lead to a positive surface density slope that can suppress type I migration \citep{Ogihara2015b}.

%FFFFFFFFFFFFFF
\begin{figure*}
    \centering
    \includegraphics[trim={1.5cm 0cm 2cm 0cm},clip,width=\textwidth]{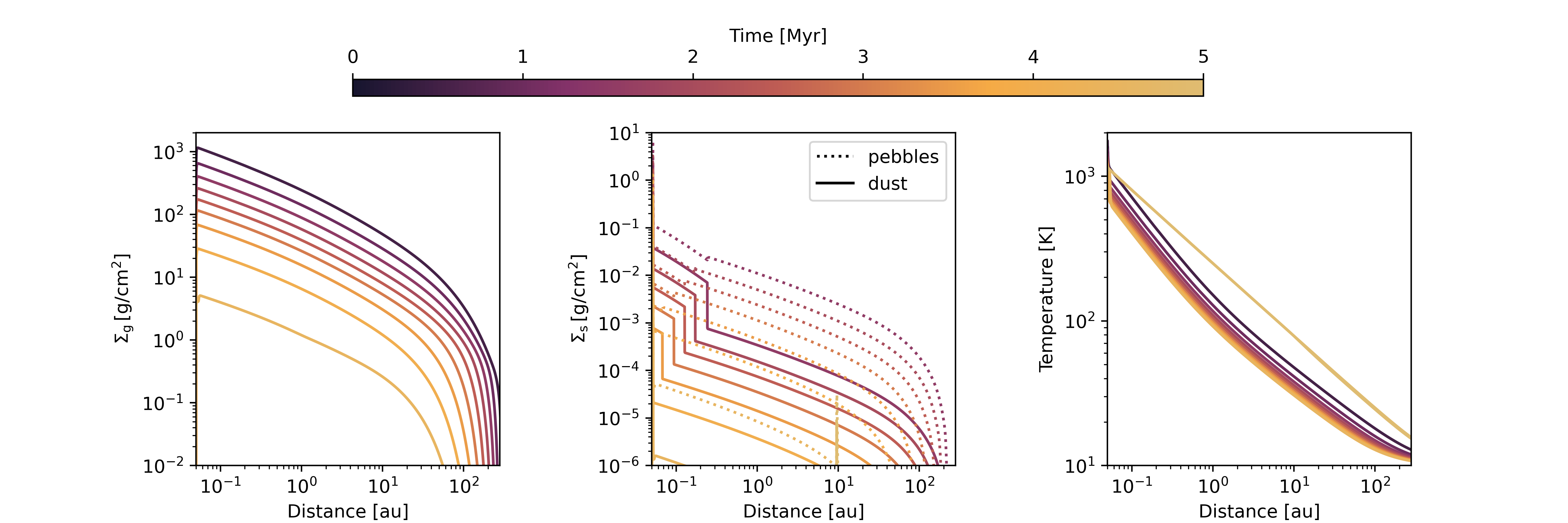}
    \caption{Time evolution of an exemplary disc evolution (see Table~\ref{tab:exemplary_disc} for the parameter choice). The left panel shows the evolution of the gas surface density. The evolution of the pebble and dust surface density evolution is shown in the middle panel and the right panel shows the time evolution of the midplane temperature. Note that the sudden change in dust surface density is related to the change from drift limited size (outer disc) to fragmentation limited size (inner disc).}
    \label{fig:exemplary_disc_evo}
\end{figure*}
%FFFFFFFFFFFFFF

% - - - - - - - 
\subsubsection{Accretion layer} \label{subsubsec:layered_accretion}
% - - - - - - - 
We assume accretion to occur in an active layer at a vertical height $z_\mathrm{active}$
\begin{equation} \label{eq:sigma_active}
    \Sigma_\mathrm{active} = \int_{z_\mathrm{active}}^{\infty} \rho(z)dz.
\end{equation}
The radial mass flow resulting from the wind torque is given by Eq.~34 in \cite{Suzuki2016}.
\begin{equation}
    \dot{M}_{r,\phi z}(r) = -\frac{4\pi}{\Omega}r\overline{\alpha_{\phi z}}(\rho c_s^2)_\mathrm{mid},
\end{equation}
with the accretion velocity $v_\mathrm{acc,active} = \dot{M}_\mathrm{r,\phi z}/(2\pi r\Sigma_\mathrm{active})$. Following \cite{Nelson2023}, we define the specific torque exerted by the wind onto the active layer
\begin{equation} \label{eq:specific_mhd_torque}
    \Lambda_\mathrm{\phi z} = \frac{\mathrm{d}\Gamma_\mathrm{\phi z}}{\mathrm{d}m} = \frac{1}{2}r\Omega v_\mathrm{acc,active} = \frac{1}{2}r\Omega \left( \frac{\dot{M}_\mathrm{r,\phi z}}{2\pi r \Sigma_\mathrm{active}} \right) = \frac{\Omega}{4\pi}\frac{\dot{M}_\mathrm{r,\phi z}}{\Sigma_\mathrm{active}}.
\end{equation}
The thickness of the active layer depends on the ionization rate (i.e. X-ray, cosmic ray and far ultraviolet (FUV) radiation) and recombination rate. We treat the layer thickness either as a constant parameter $\Sigma_\mathrm{active}$ or assume accretion in the layer to occur at a fraction of sonic velocity
\begin{equation}
    \Sigma_\mathrm{active}(f_\mathrm{s} \times c_\mathrm{s}) = \frac{\dot{M}_\mathrm{r,\phi z}}{2\pi r f_\mathrm{s}c_\mathrm{s}}.
\end{equation}
The specific torque of the latter shows a much weaker scaling with orbital distance $\Lambda_\mathrm{\phi z} \propto r^{-3/4}$ than compared to the constant accretion layer thickness $\Lambda_\mathrm{\phi z} \propto r^{-7/4}$.\footnote{For these scalings we assumed $\Sigma_\mathrm{g} \propto r^{-1}$ and a temperature profile of a passively irradiated disc $T_\mathrm{mid} \propto r^{-1/2}$.} We point out that since the specific torque depends only on $c_\mathrm{s}$ for $\Sigma_\mathrm{active}(f_\mathrm{s}\times c_\mathrm{s})$ it will show very little temporal evolution, however for constant $\Sigma_\mathrm{active}$ the specific torque will decrease along with the accretion rate.

% - - - - - - - 
\subsubsection{Thermal structure} \label{subsubsec:thermal_structure}
% - - - - - - - 
The 1D vertically integrated model makes so far no assumption to where accretion is occurring. \cite{Mori2021} developed a model for the thermal structure in a disc, where accretion and consequently accretion heating is occurring in an accretion layer. The contribution to the heating of the disc midplane is given as
\begin{equation}
    T_\mathrm{acc,mid} = \left[ \left( \frac{3F_\mathrm{rad}}{8\sigma_\mathrm{SB}} \right) \left( \kappa_\mathrm{R} \Sigma_\mathrm{heat} + \frac{1}{\sqrt{3}}\right) \right]^{1/4},
\end{equation}
with $\kappa_\mathrm{R}$ being the Rosseland-mean opacity and $\Sigma_\mathrm{heat}$ being the integrated density from infinity to the bottom of the heating region, which is equivalent to $\Sigma_\mathrm{active}$ \citep{Mori2021}. Gas opacities from \cite{Malygin2014} and dust opacities from \cite{Semenov2003} are used to calculate $\kappa_\mathrm{R}$, following \cite{Marleau2017,Marleau2019}.
The midplane temperature is given by
\begin{equation}
    T_\mathrm{mid} = ( T_\mathrm{acc,mid}^4 + T_\mathrm{ext}^4 )^{1/4},
\end{equation}
with $T_\mathrm{ext}$ being contributions due to external irradiation, such as irradiation from the star and surrounding molecular cloud \cite[see Eq.~3, Eq.~4 and Eq.~5 in][]{Weder2023}.

% - - - - - - - 
\subsubsection{Photoevaporative winds} \label{subsubsec:photoevaporative wind}
% - - - - - - - 
Our model also includes outflows from internal photoevaporation by extreme ultraviolet (EUV) radiation from the host star using a diffuse stellar irradiation model \cite[see Appendix A in][]{Alexander2007}. We further account for shielding of the EUV radiation by the merging MHD wind \cite[see][for details on the shielding mechanism]{Weder2023}. We also include external photoevaporation from far ultraviolet (FUV) irradiation from surrounding stars using the FRIED grid v2 \citep{haworth_fried_2023}. For details on the implementation we refer to \cite{Weder2023,Weder2025b}. We here adopt a low value for the ambient FUV field strength ($\mathrm{10\,G_0}$) and polycyclic aromatic hydrocarbon (PAH) abundance of $0.1$ such that the disc evolution is not strongly influenced by external photoevaporation.

% - - - - - - - 
\subsubsection{Dust evolution} \label{subsubsec:dust_evo}
% - - - - - - - 
For the dust evolution we follow the two population model of dust growth from \cite{Birnstiel2012}, where both dust and pebble surface density evolution are considered ($\Sigma_\mathrm{dust}(r,t)$ and $\Sigma_\mathrm{pebble}(r,t)$). The evolution of the combined solid disc $\Sigma_\mathrm{s}=\Sigma_\mathrm{dust}+\Sigma_\mathrm{pebble}$ is calculated as a combined advection and diffusion equation. For details on the implementation we refer to \cite{Voelkel2020} and \cite{Burn2022}. The surface density profiles of the two population is given at each time step as $\Sigma_\mathrm{dust}(r)=\Sigma_\mathrm{s}(r)(1-f_m(r))$ and $\Sigma_\mathrm{pebble}(r)=\Sigma_\mathrm{s}(r)f_m(r)$ with $f_m(r)$ being dependent on whether drift limited or fragmentation limited regime prevails. The grain sizes in the two populations are given through 
\begin{eqnarray}
    a_0&=&10^{-5}\mathrm{cm} \\
    a_\mathrm{max}(t)&=&\min\left[a_\mathrm{drift},a_\mathrm{frag},a_0\cdot e^{t f_\mathrm{\small{D/G}}\Omega}\right],
\end{eqnarray}
where the latter includes grain growth, drift and fragmentation at a threshold $v_\mathrm{frag}=10\,\mathrm{m/s}$, neglecting compositional dependencies. The fragmentation limit is correlated with the turbulent viscosity $a_\mathrm{frag} \propto \Sigma_\mathrm{g}/\overline{\alpha_{r\phi}}$ whereas the drift limited $a_\mathrm{drift} \propto \Sigma_\mathrm{s}$ \citep{Zagaria2022}.

% - - - - - - - 
\subsubsection{Exemplary disc evolution} \label{subsubsec:exempl_disc_evo}
% - - - - - - - 
Figure~\ref{fig:exemplary_disc_evo} shows the time evolution of an exemplary disc. The choices for parameters are listed in Table~\ref{tab:exemplary_disc}. The disc has a NIR-lifetime of $\simeq 4.7\,\mathrm{Myr}$ which is in line with expected lifetimes of protoplanetary discs \cite[see][for details on the disc dispersal condition]{Weder2023}.

Note the rapid increase in dust surface density that corresponds to the transition from drift- to fragmentation limited regime. The here presented case is similar to the hybrid case discussed in \cite{Zagaria2022}.

The heating of the disc midplane appears to be very ineffective and the midplane temperature is similar to a passively irradiated disc \cite[see also Fig.~2 in][]{Mori2025}.

%TTTTTTTTTTTTTT
\begin{table}[!ht]
    \begin{center}
    \caption{Parameter choices for the exemplary case.}
    \label{tab:exemplary_disc}
    \begin{tabular}{l c l}
    \hline \hline
    Parameter & Symbol & Value \\ \hline
    Stellar mass & $M_\star$ & $1\,\mathrm{M_\odot}$ \\
    Disc mass & $M_\mathrm{disc}$ & $0.1\,\mathrm{M_\odot}$ \\
    Inner radius & $R_\mathrm{in}$ & $0.05\,\mathrm{au}$ \\
    Characteristic radius & $R_\mathrm{char}$ & $100\,\mathrm{au}$ \\
    Turbulent viscosity & $\overline{\alpha_{r\phi}}$ & $10^{-4}$ \\
    Wind stress & $\overline{\alpha_{\phi z}}$ & $2 \cdot 10^{-3}$ \\
    Accretion layer & $\Sigma_\mathrm{active}$ & $0.1\,\mathrm{g/cm^2}$ \\
    Dust to gas ratio & $f_\mathrm{D/G}$ & $0.014$ \\
    \hline
    \end{tabular}
    \end{center}
\end{table}
%TTTTTTTTTTTTTT

%---------------
\subsection{Planetary migration model} \label{subsec:migration_model}
\begin{figure*}[htbp]
  \centering
  \begin{subfigure}{0.48\textwidth}
    \centering
    \includegraphics[width=\textwidth]{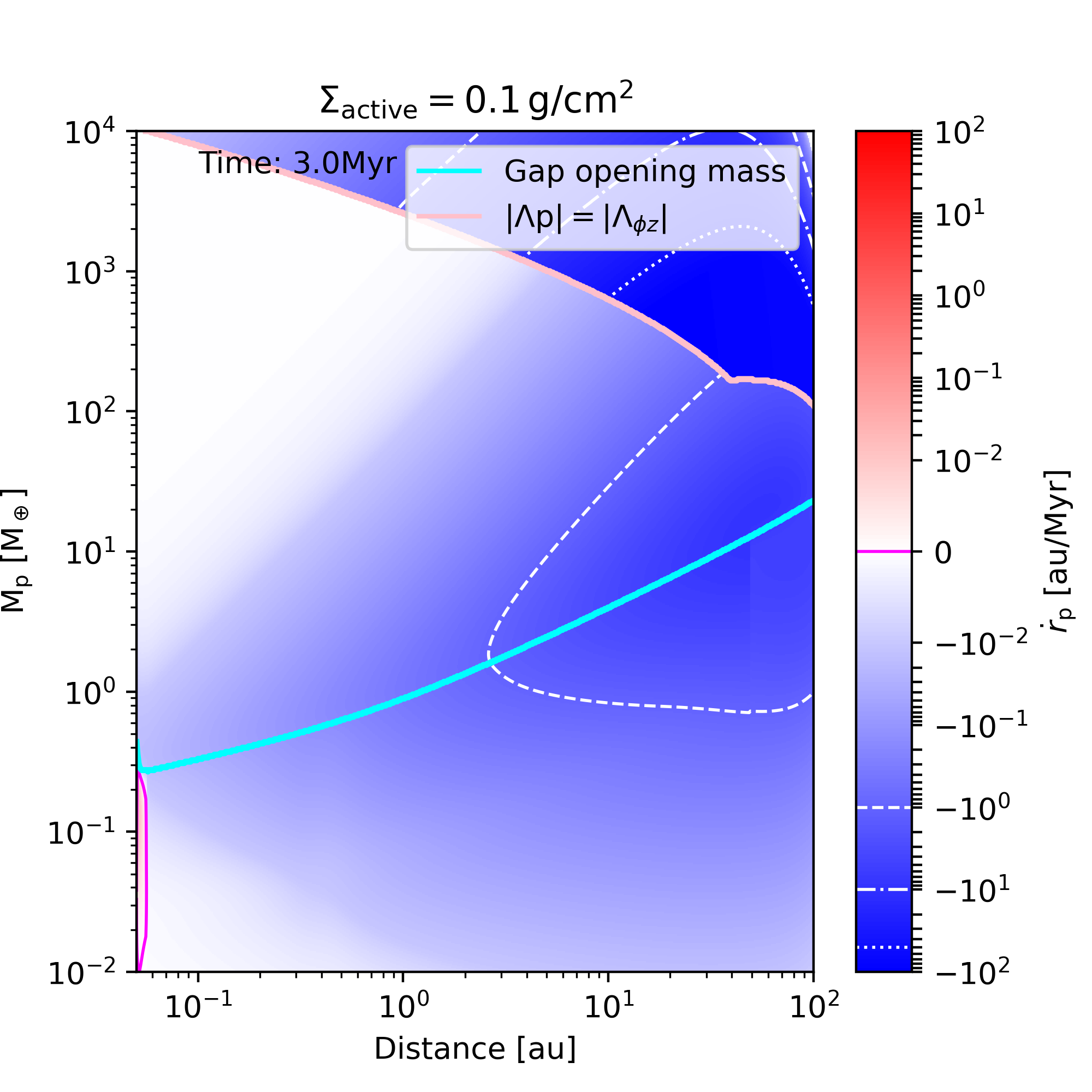}
  \end{subfigure}
  \hfill
  \begin{subfigure}{0.48\textwidth}
    \centering
    \includegraphics[width=\textwidth]{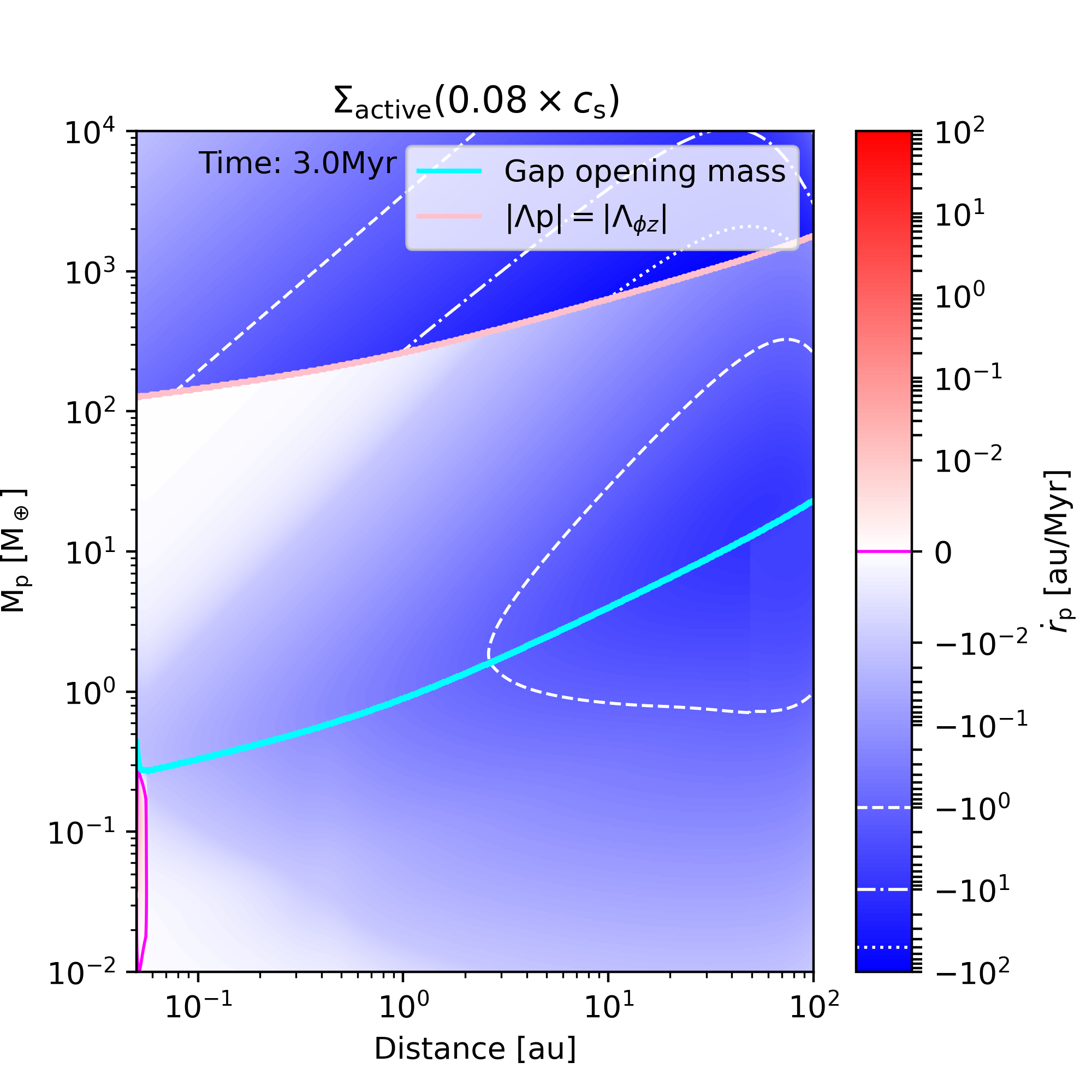}
  \end{subfigure}
    \caption{Migration rate as a function of orbital distance and planetary mass for the exemplary disc evolution at 3~Myrs (see Sect.~\ref{subsubsec:exempl_disc_evo}). The cyan line marks the transition from Type I to Type II migration through gap opening. The pink line denotes the transition into the wind-driven Type II migration regime. The left panel shows the migration map for $\Sigma_\mathrm{active}=0.1\,\mathrm{g/cm^2}$ and the right panel shows the migration map for $\Sigma_\mathrm{active}(0.08\times c_\mathrm{s})$.}
    \label{fig:migration_regimes}
\end{figure*}

%---------------
A schematic view of the migration regimes is given in Fig.~\ref{fig:concept_overview} and a migration map showing all different migration regimes is shown in Fig.~\ref{fig:migration_regimes}.

%FFFFFFFFFFFFFF
%FFFFFFFFFFFFFF

%FFFFFFFFFFFFFF
\begin{figure}[htbp]
    \centering
    \includegraphics[width=\linewidth]{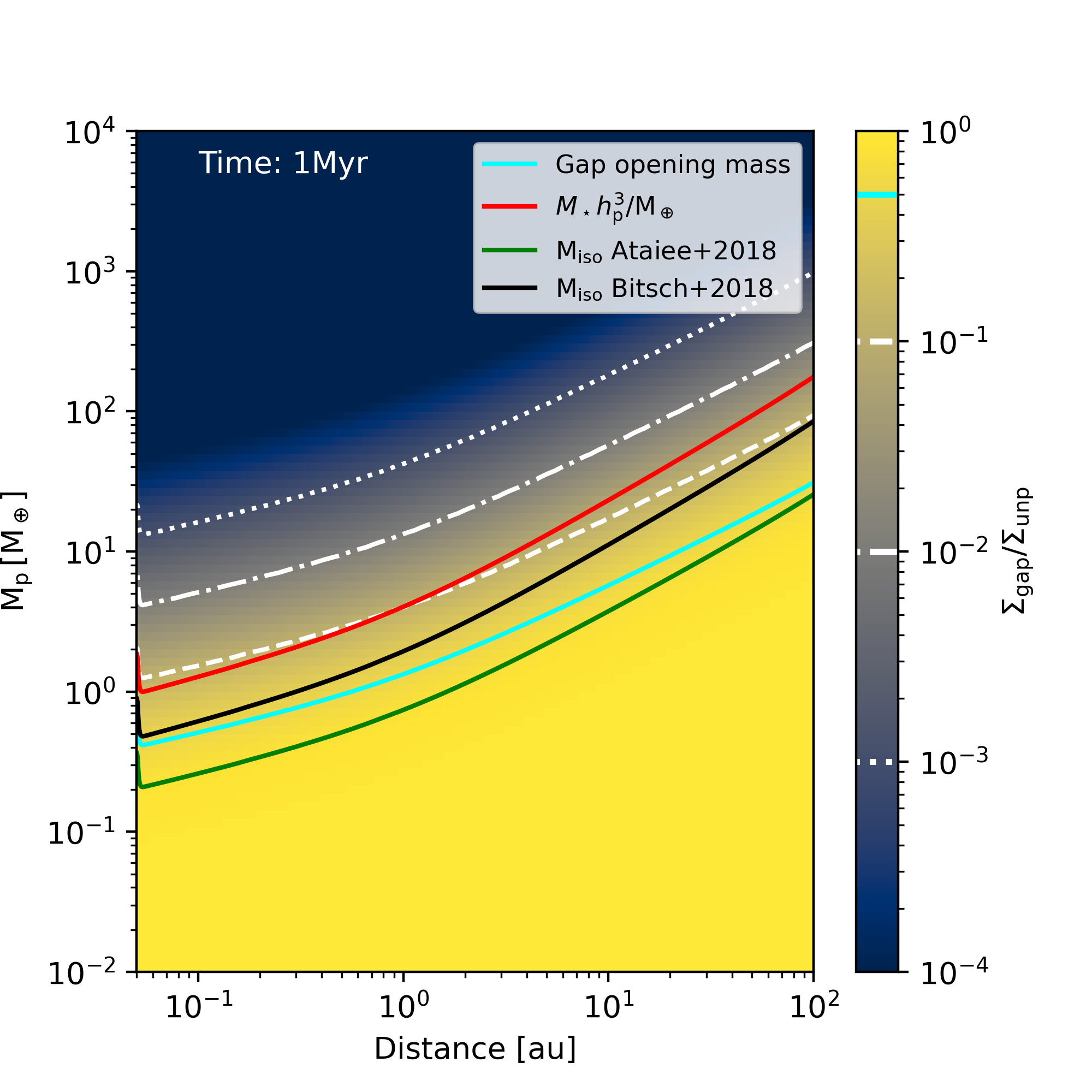}
    \caption{Map showing the reduced surface density in the gap as a function of planetary mass for the exemplary disc evolution at 1~Myr (see Sect.~\ref{subsubsec:exempl_disc_evo}), calculated using Eq.~\ref{eq:kanag_gap_depth}. The cyan line corresponds to the gap opening mass with the criterion $\Sigma_\mathrm{gap}/\Sigma_\mathrm{unp} = 0.5$. We show pebble isolation mass formulas from \cite{Bitsch2018} and \cite{Ataiee2018} in green and black and the red line corresponds to the thermal mass criterion.}
    \label{fig:gap_depth}
\end{figure}
%FFFFFFFFFFFFFF

% - - - - - - - 
\subsubsection{Type I migration} \label{subsubsec:type_I_mig}
% - - - - - - - 
The migration of low-mass planets that are still embedded in the disc is not expected to be influenced by the presence of an accreting layer \citep{McNally2020}. We therefore use the Type I migration prescription from \cite{Paardekooper2011}. However, since the midplane has a low turbulent viscosity, we include the dynamical corotation torque using the memory timescale approach in \citep{Weder2025}. Note that Fig.~\ref{fig:migration_regimes} does not include the effect of the dynamical corotation torque on Type I migration rates as it is partly dependent on the planet's initial location.

% - - - - - - - 
\subsubsection{Gap opening} \label{subsubsec:gap_opening}
% - - - - - - - 
With increasing mass, the planet starts to perturb the gas surface density, eventually opening a deep gap. While gap formation in viscous discs has been studied extensively \citep{Crida2006,Crida2007,Kanagawa2016,Kangawa2018} gap formation in low viscosity and MHD wind-driven discs remains unclear. \cite{Elbakyan2022} studied gap formation in discs with laminar flows driven by magnetized winds, using 2D simulations with a prescribed torque. They found that gap opening is generally determined by the residual turbulence and proposed a simple modification to the criterion from \cite{Crida2006}. \cite{AoyamaBai2023} and \cite{Wafflard-Fernandez2023} both conducted 3D global non-ideal MHD simulations including gap opening planets. \cite{AoyamaBai2023} found gap shapes similar to inviscid discs, however much deeper due to the inhomogeneous wind torque as a result to the magnetic flux concentration in the gap. This seems to be in contrast to \cite{Wafflard-Fernandez2023} where they found that higher initial magnetisation (lower $\beta$ and hence higher $\overline{\alpha_\mathrm{\phi z}}$) leads to a higher gap opening mass (see their Fig.~5), similar to \cite{Elbakyan2022}.
Furthermore, the opening of deep gaps leads to steep gap edges that are Rossby-wave unstable and small scale vortices appear to diffuse the gap edges \citep[e.g.][]{McNally2019}. 

Given these uncertainties, we here use the criteria from \cite{Kanagawa2016}, which does not include the dependence on the magnetic field strength. A major difference between the gap criteria from \cite{Kanagawa2016} and \cite{Crida2006} is the inclusion of the so called thermal criterion $q_\mathrm{p} \gtrsim h^3$ \citep{LinPapaloizou1993}. In absence of accretion heating processes, this is in fact similar to the viscous condition for gap formation (see Fig.~\ref{fig:gap_depth}). The surface density inside the gap is given by 
\begin{equation}\label{eq:kanag_gap_depth}
    \frac{\Sigma_\mathrm{gap}}{\Sigma_\mathrm{unp}} = \frac{1}{1+0.04K},
\end{equation}
where
\begin{equation}
    K=\left( \frac{M_\mathrm{p}}{M_\star} \right)^{2}\left( \frac{H_\mathrm{p}}{r_\mathrm{p}} \right)^{-5}\overline{\alpha_\mathrm{r\phi}}^{-1} = q^{2}h_\mathrm{p}^{-5}\overline{\alpha_\mathrm{r\phi}}^{-1}.
\end{equation}
Gap opening is considered if the perturbed surface density reaches $\Sigma_\mathrm{unp}/2$. Figure~\ref{fig:gap_depth} shows the gap depth as a function of orbital distance and planetary mass for the exemplary disc discussed in Sect.~\ref{subsubsec:exempl_disc_evo} at 1~Myr.

% - - - - - - - 
\subsubsection{Type II migration} \label{subsubsec:type_II_mig}
% - - - - - - - 
As the planet opens a gap, it eventually enters the Type II migration regime. In contrast to the Type I regime, the Type II regime is expected to differ significantly from the classic viscous model due to the fast accretion flow and or different gap shapes. \citet{AoyamaBai2023} find Type II migration to be generally inwards, potentially affected by structures in the disc. In contrast, \citet{Wafflard-Fernandez2023} find sustained outward migration for Jovian planets due to asymmetric gaps and stochastic but net slow outward migration for low-mass planets. \cite{Lega2022} found two distinct migration regimes depending on whether the planet torque onto the disc $|\Lambda_\mathrm{p}|$ exceeds the magnetic torque $|\Lambda_\mathrm{\phi z}|$, which can lead to rapid inwards migration of giant planets - a potential explanation for the existence of hot and warm Jupiters. We here investigate the implication of this mechanism on a global scale, following the analytical description given in \cite{Nelson2023}, where the torque density described below is used to distinguish between these two regimes.

The angular momentum exchange between planet and disc can be expressed using torque densities $\Lambda_\mathrm{p}=\mathrm{d}\Gamma_\mathrm{p}/\mathrm{d}m$ \citep{AngeloLubow2010}
\begin{equation}
    \Lambda_\mathrm{p} = \mathcal{F}(x,\beta,\zeta) \Omega_\mathrm{p}^2 r_\mathrm{p}^2 q^2 \left( \frac{r_\mathrm{p}}{\Delta_\mathrm{p}} \right)^{4},
\end{equation}
with $\beta=-\mathrm{d}\ln\Sigma_\mathrm{g}/\mathrm{d}\ln r$ and $\zeta=-\mathrm{d}\ln T/\mathrm{d}\ln r$. $\Delta_\mathrm{p}$ corresponds to the the maximum of the vertical scale height and the Hill radius (i.e. $\Delta_\mathrm{p}=\max(H_\mathrm{p},R_\mathrm{Hill})$). The function $\mathcal{F}$ is defined as
\begin{equation}
    \mathcal{F}(x,\beta,\zeta) = \left[ p_1 e^{-(x+p_2)^2/p_3^2} + p_4 e^{-(x-p_5)^2/p_6^2} \right] \cdot \tanh(p_7-p_8x),
\end{equation}
with $x=(r-r_\mathrm{p})/\Delta_\mathrm{p}$ and $p_i$ being parameters obtained from fits \cite[see Table~1 in][]{AngeloLubow2010}.

\paragraph{Viscosity dominated regime $|\Lambda_\mathrm{p}|<|\Lambda_\mathrm{\phi z}|$}
\cite{Kimmig2020} investigated the migration behaviour of giant planets in a disc with a laminar MHD wind driven accretion flow using 2D hydrodynamical simulations. They found that replenishment of the co-orbital region can lead to a Type-III-like outward migration. Their simulations suggest that this is the case if the timescale for the material to pass the librating region is sufficiently small in comparison to the libration timescale ($\tau_\mathrm{pass}/\tau_\mathrm{lib} \lesssim 10$). However, in order to have outward migration, very high accretion rates through the disc have to be reached (i.e. $\dot{M}_\mathrm{r,\phi z} \simeq 3\cdot 10^{-6} \,\mathrm{M_\odot/yr}$), which is much higher than what we have in our simulations and we therefore neglect this process.

In this regime, the negative wind torque exceeds the positive planet torque and the gas flows unimpeded through the gap. We follow the approach from \cite{Kangawa2018}, where the torque onto the planet is related to $\Sigma_\mathrm{gap}$. The migration rate is then given as
\begin{equation}
    \dot{r}_\mathrm{p} = -150 \frac{\Sigma_\mathrm{g}r_\mathrm{p}}{M_\mathrm{p}}\overline{\alpha_\mathrm{r\phi}} h^{3} \Omega_\mathrm{p}.
\end{equation}
Since $\overline{\alpha_\mathrm{r\phi}}$ is low, gaps are deep and type II migration in this regime is slow compared to viscous discs.

\paragraph{Wind-driven regime $|\Lambda_\mathrm{p}|>|\Lambda_\mathrm{\phi z}|$}
Once the planet torque exceeds the magnetic torque, the gas is halted and starts to pile up at the outer edge of the gap, leading to fast inwards migration \citep{Lega2022}. This so called wind-driven migration occurs at the rate the active layer can replenish the outer gap edge
\begin{equation}\label{eq:wind_driven_disc_dominated}
    \dot{r}_\mathrm{p} = \frac{\dot{M}_\mathrm{r,\phi z}}{2\pi r_\mathrm{p}\Sigma_\mathrm{g}}\,.
\end{equation}
However, note that the planet only is expected to migrate if the the surface density of the disc is large enough. In the limit of $\Sigma_\mathrm{g} \rightarrow 0$ the migration rate in Eq.~\ref{eq:wind_driven_disc_dominated} goes to infinity. This is because we assume that the torque exerted onto a planet by the surface density of the outer gap is enough to push the planet. If one assumes that the planet feels the torque exerted by the mass flow in the outer gap $\Gamma_\mathrm{p}=r_\mathrm{p}^2\Omega\dot{M}_\mathrm{r,\phi z}$, we get the following expression for the migration velocity
\begin{equation}\label{eq:wind_dirven_planet_dominated}
    \dot{r}_\mathrm{p} = \frac{2\Gamma_\mathrm{p}}{M_\mathrm{p}r_\mathrm{p}\Omega} = \frac{2r_\mathrm{p}\dot{M}_\mathrm{r,\phi z}}{M_\mathrm{p}}
\end{equation}
The migration rate of the planet is set by the minimum of Eqs.~\ref{eq:wind_driven_disc_dominated} and \ref{eq:wind_dirven_planet_dominated}, as it cannot migrate faster than the outer gap is replenished, but also does only feel the torque of the gap edge. The migration rate can thus be written as
\begin{equation} \label{eq:wind_driven_full}
    \dot{r}_\mathrm{p} = \dot{M}_\mathrm{r,\phi z} \cdot \min \left( \frac{1}{2\pi r_\mathrm{p}\Sigma_\mathrm{g}}, \frac{2r_\mathrm{p}}{M_\mathrm{p}} \right)
\end{equation}
Note that in the limit of a purely viscously evolving disc, the steady state accretion rate is given by $\dot{M}_\mathrm{acc} \simeq 3\pi \nu \Sigma_\mathrm{g}$. This naturally results in the classical migration rates for the disc- and planet dominated case \cite[see Eq.~12 in ][]{Paardekooper_PPVII_2023}.

Fig.~\ref{fig:migration_regimes} shows transition to the wind driven regime at $|\Lambda_\mathrm{p}| = |\Lambda_\mathrm{\phi z}|$ for both a constant $\Sigma_\mathrm{active}$ (left panel) and $\Sigma_\mathrm{active}(f_\mathrm{s}\times c_\mathrm{s})$ (right panel). At high planetary mass $R_\mathrm{Hill}>H_\mathrm{p}$ and thus we can write the scaling of the planet torque as $\Lambda_\mathrm{p} \propto r^{-1} M_\mathrm{p}^{2/3}$. Recalling the scalings of $\Lambda_\mathrm{\phi z}$ derived in Sect.~\ref{subsubsec:layered_accretion} shows indeed that the critical mass scales with $\propto r^{0.4}$ for $\Sigma_\mathrm{active}(f_\mathrm{s}\times c_\mathrm{s})$, whereas for a constant $\Sigma_\mathrm{active}$ it is a decreasing function of orbital distance $\propto r^{-1.1}$ and time $\propto \dot{M}_\mathrm{r,\phi z}$.\footnote{Note that the scaling with $r$ is less precise for constant $\Sigma_\mathrm{active}$ as the dependencies on $\dot{M}_\mathrm{\phi z}$ and $\Sigma_\mathrm{g}$ do not cancel out when comparing $\Lambda_\mathrm{p}$ and $\Lambda_\mathrm{\phi z}$ and hence the assumptions made in Sect.~\ref{subsubsec:layered_accretion} may not hold after some time.}

%---------------
\subsection{Planetary accretion model} \label{subsec:planet_growth}
%---------------
The starting point in our simulations are lunar mass embryos ($10^{-2}\,\mathrm{M_\oplus}$) that are inserted in the disc and accrete pebbles and gas. Planetary embryos are believed to form as a result of pressure maxima \citep{Xu2022,Zhao2025} or streaming instability \citep{Youdin2005}. Spontaneous and stochastic magnetic flux concentration naturally leads to the formation of disc substructures \cite[e.g.][]{Riols2019,Riols2020}, which poses a potential starting point of forming planetary embryos.

% - - - - - - - 
\subsubsection{Pebble accretion} \label{subsubsec:pebble_accretion}
% - - - - - - - 
We simplify the solid accretion process by assuming that it occurs via pebbles  \citep[see][for a discussion of the importance of the different solid accretion mechanisms]{MordasiniBurn2024} and follow \cite{Ormel2017} to model the accretion of pebbles. The vertical distribution of particles with Stokes number $\mathrm{St}$ in the gas disc is dependent on the turbulent viscosity $H_\mathrm{peb} = H \sqrt{\overline{\alpha_\mathrm{r\phi}}/(\overline{\alpha_\mathrm{r\phi}}+\mathrm{St})}$. If pebbles reside in a thin layer compared to the accretion impact parameter ($H_\mathrm{peb}\ll b_\mathrm{acc}$) the accretion rate is given as
\begin{equation}
    \dot{M}_\mathrm{2D} = 2b_\mathrm{acc}v_\mathrm{enc}(b_\mathrm{acc})\Sigma_\mathrm{peb}.
\end{equation}
Here $b_\mathrm{acc}$ corresponds to the largest impact parameter that fulfils both $t_\mathrm{settle}<t_\mathrm{enc}$ and $t_\mathrm{stop}<t_\mathrm{enc}$ \citep[see][]{Ormel2017} and $v_\mathrm{enc}(b_\mathrm{acc})$ is the relative encounter velocity between planet and pebble \citep[see Eq.~7.6 in][]{Ormel2017}. Since we here focus on discs with low turbulent viscosity, we expect generally $H_\mathrm{peb} \ll b_\mathrm{acc}$ to be true. However, for small impact parameters (i.e. small planet mass) it can be that $H_\mathrm{peb} \gg b_\mathrm{acc}$ and the accretion rate is then given as
\begin{equation}
    \dot{M}_\mathrm{peb} = \dot{M}_\mathrm{2D} \frac{b_\mathrm{acc}}{b_\mathrm{acc}+\sqrt{8/\pi}H_\mathrm{peb}}.
\end{equation}

Pebble accretion relies on a sustained influx of pebbles from the outer disc. However, as the planet grows, it starts to influence the gas surface density and create a gap (see Sect.~\ref{subsubsec:gap_opening}). The outer edge of the gap corresponds to a pressure bump that acts as a pebble trap. Once this so called pebble isolation mass is reached, pebble accretion is shut off. We use the analytical prescription for the pebble isolation mass by \cite{Ataiee2018}:
\begin{equation}
    \left( \frac{M_\mathrm{p}}{M_\star} \right)_\mathrm{iso} \approx h^3 \sqrt{37.3\overline{\alpha_\mathrm{r\phi}}+0.01} \times \left[ 1 + 0.2\left( \frac{\sqrt{\overline{\alpha_\mathrm{r\phi}}}}{h} \sqrt{\frac{1}{\mathrm{St}^2}+4} \right)^{0.7} \right].
\end{equation}

% - - - - - - - 
\subsubsection{Gas accretion} \label{subsubsec:gas_accretion_model}
% - - - - - - - 
In the attached phase, the envelope is in equilibrium with the gas disc and the accretion rate is limited by the planets ability to radiate away the liberated potential energy. Here we calculate the gas accretion rate by solving the internal structure equations \cite[see Sect.~4.1.1 in][]{Emsenhuber_NGPPSI_2021}. Simultaneously we calculate the maximum gas accretion rate following the calculations from \cite{Choksi2023} who derived scaling laws with the thermal mass $q_\mathrm{th} \equiv (M_\mathrm{p}/M_\star)(H_\mathrm{p}/r_\mathrm{p})^{-3}$:
\begin{equation}
    \dot{M}_\mathrm{env,max} = C_i q_\mathrm{th,i}^{a_i} h_\mathrm{p}^{3} r_\mathrm{p}^{3} \Omega_\mathrm{p}\rho_\mathrm{gas,p},
\end{equation}
with $q_\mathrm{th,1} \lesssim 0.3 \lesssim q_\mathrm{th,2} \lesssim 10 \lesssim q_\mathrm{th,3}$ with corresponding $C_1=3.5,C_2=4/3,C_3=9/3^{2/3}$ and $a_1=2,a_2=1,a_3=2/3$. Note that the gas density corresponds to the density at the planets location (inside the gap) $\rho_\mathrm{gas,p} = \Sigma_\mathrm{gap}/(\sqrt{2\pi}H_\mathrm{p})$. The envelope accretion rate is then given by the minimum of the two rates.

\cite{Nelson2023} found that the accretion rate onto the planet is related to the planets ability to block the flow. In the viscosity dominated regime ($|\Lambda_\mathrm{p}|<|\Lambda_\mathrm{\phi z}|$) the planet continues to accrete at high rates, as long as the unimpeded accretion flow reaches the planet. We include this effect in an approximate way by adding the fraction of the flow that goes through the planets Hill sphere to the planet
\begin{equation}
    \dot{M}_\mathrm{env} = \min \left(\dot{M}_\mathrm{env,attached},\dot{M}_\mathrm{env,max} + \dot{M}_\mathrm{r,\phi z} \frac{2R_\mathrm{Hill}}{2\pi r_\mathrm{p}} \right).
\end{equation}
When the planet enters the wind-driven migration regime ($|\Lambda_\mathrm{p}|>|\Lambda_\mathrm{\phi z}|$), the accretion flow is blocked and accretion from the wind-driven layer (term proportional to $\dot{M}_\mathrm{r,\phi z})$ is gradually shut down by evaluating at what distance from the planet its torque intercepts the accretion flow $R_\mathrm{intercept}$ and reduce the accretion rate by a factor $f_\mathrm{red}=\max[3/2-R_\mathrm{intercept}/(2R_\mathrm{Hill}),0]$ that is linearly decreasing out to a distance of $3\,R_\mathrm{Hill}$, which corresponds to the upper limit where the planets accrete from \citep{Machida2010}.

%---------------
\subsection{Exemplary case: formation tracks} \label{subsec:example_formation_tracks}
%---------------

%FFFFFFFFFFFFFF
\begin{figure*}[htb]
    \centering
    \includegraphics[width=\textwidth]{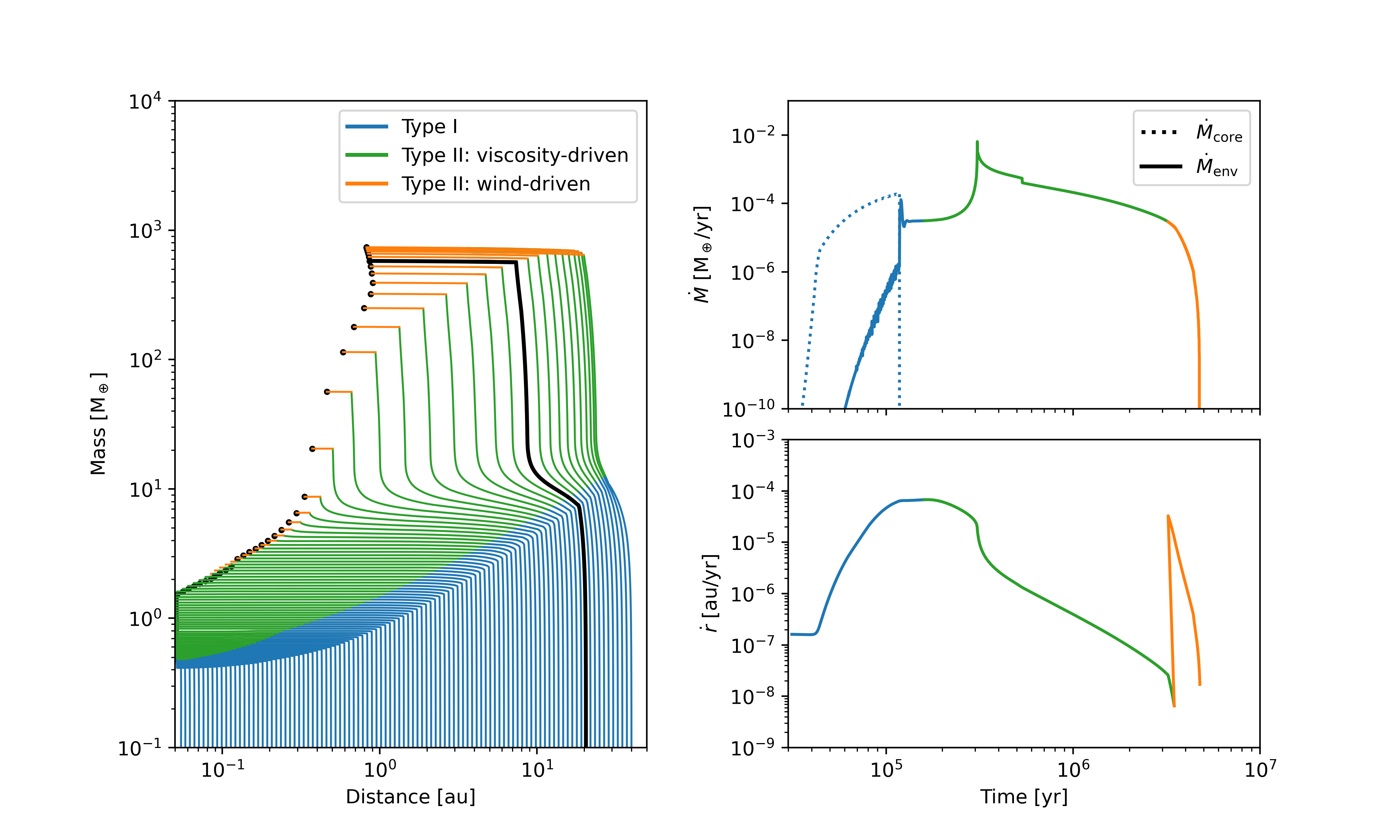}
    \caption{Formation tracks compiled from 100 individual single embryo simulations ($M_\mathrm{emb}=10^{-2}\,\mathrm{M_\oplus}$) with varying initial locations $r_\mathrm{start}$ spaced uniform in log between $0.05\,\mathrm{au}$ and $40\,\mathrm{au}$ for our exemplary disc discussed in Section~\ref{subsubsec:exempl_disc_evo}. Tracks are coloured by migration regime. Time evolution of the planet's accretion rate (core and envelope) and migration rate are shown as an example for a planet with $r_\mathrm{start}\simeq20\,\mathrm{au}$.}
    \label{fig:example_formation_tracks}
\end{figure*}
%FFFFFFFFFFFFFF

Fig.~\ref{fig:example_formation_tracks} shows the formation tracks for lunar mass embryos ($10^{-2}\,\mathrm{M_\oplus}$) inserted at $t=0$ with varying initial location in the (same) exemplary disc (see Section~\ref{subsubsec:exempl_disc_evo} and Table~\ref{tab:exemplary_disc}). Each planet forms individually (no N-body interactions). Assuming embryos being able to form anywhere in the disc at $t=0$ may be a gross simplification. The focus of this work, however, lies on the interaction between the planets and the gas disc. Initially, the planets grow fast due to pebble accretion until they reach the pebble isolation mass and the core accretion rate goes to zero. The associated drop of the luminosity in the planet's envelope leads  to an increase of the gas accretion rate \citep{KesslerAlibert2023}. While the envelope is still accreting, however at a lower rate than the core accretion rate before, they rapidly migrate inwards (Type I migration). They eventually start opening a deep gap and migration slows down (Type II migration). Planets starting further out in the disc $\gtrsim10\,\mathrm{au}$ eventually go in runaway growth. The right panels in Fig.~\ref{fig:example_formation_tracks} show the growth and migration rate of a planet that starts at $\simeq20\,\mathrm{au}$. During gas runaway accretion, the migration rate starts to rapidly decrease due to the deep gap. This also eventually causes the gas accretion rate to reduce. Eventually, they enter the fast wind-driven Type II migration regime. We point out that they either enter the wind-driven regime by growing, or by stalling during growth and the decreasing $\Lambda_\mathrm{\phi z} \propto \dot{M}_\mathrm{r,\phi z}$ picks them up, which is why planets enter this regime at different masses.

%===============
\section{Single-embryo planetary population syntheses (SEPPS)} \label{sec:results}
%===============
After having discussed the model and the resulting accretion and migration regimes at an exemplary case we continue assessing the impact of different paremeterisations for $\Sigma_\mathrm{active}$ on planet formation by performing single-embryo planetary population syntheses (SEPPS). We considered discs with varying initial conditions around a solar mass star. We first present the initial conditions in Sect.~\ref{subsec:SEPPS_init_conds} and Appendix \ref{app:init_dist} and show the results in Sect.~\ref{subsec:SEPPS_results}.

%---------------
\subsection{SEPPS - initial conditions} \label{subsec:SEPPS_init_conds}
%---------------
For the distributions of the initial conditions of the discs we follow closely \cite{Weder2025b} and they are shown in Appendix \ref{app:init_dist}. Dust masses are inferred from a log-normal fit to Class 1 protoplanetary discs from \cite{Tychoniec2018} that is representative for stellar masses of $\sim0.3\,\mathrm{M_\odot}$. Assuming linear scaling with stellar mass this results in $\log_{10}(\mu/\mathrm{M_\oplus})=2.6$ and $\sigma=0.35\,\mathrm{dex}$. The characteristic radius is calculated using the relation $R_\mathrm{char}=70\,\mathrm{au}\cdot\left[ M_\mathrm{dust}/100\mathrm{M_\oplus} \right]^{0.25}$ with a $1\,\mathrm{dex}$ spread \citep{Tobin2020}. We use a normal distributed metallicity inferred from observations \citep[$\mu=-0.02$ and $\sigma=0.22$,][]{Santos2005} to calculate the dust-to-gas ratios using the relation $f_\mathrm{D/G}=f_\mathrm{D/G,\odot}10^{\mathrm{[Fe/H]}}$. The gas masses are then given by $M_\mathrm{dust}/f_\mathrm{D/G}$. The strength of angular momentum transport is derived assuming a log-uniform distribution for the accretion timescale $\log_{10}(\tau_\mathrm{acc})=\mathcal{U}(-0.6,0.4)$. Evaluating the expression for the accretion timescale $\tau_\mathrm{acc}=M_\mathrm{disc}/\dot{M}_\mathrm{acc}$ at the characteristic radius results in a formulation for the wind stress $\overline{\alpha_\mathrm{\phi z}}$ as a function of the accretion timescale $\tau_\mathrm{acc}$:
\begin{equation}
    \overline{\alpha_\mathrm{\phi z}} = \frac{M_\mathrm{disc}}{2\sqrt{2\pi}r^2h\Omega \Sigma \tau_\mathrm{acc}},
\end{equation}
with $h$, $\Omega$, and $\Sigma$ being evaluated at the characteristic radius $R_\mathrm{char}$. For the external irradiation we again adopt a low value of $10\,\mathrm{G_0}$ for the ambient FUV field strength as it is not our goal to investigate the impact of strong external photoevaporation. 

The goal is to test different values and parameterisations of the accretion layer thickness $\Sigma_\mathrm{active}$. We adopt either a constant value of $\Sigma_\mathrm{active}$ or assume accretion to occur at a fraction of the sonic velocity $f_\mathrm{s} \cdot c_\mathrm{s}$ (see Sect.~\ref{subsubsec:layered_accretion}).

Lunar mass embryos are inserted randomly in log uniform $\log_{10}(r_\mathrm{start}) = \mathcal{U}(\log_{10}(R_\mathrm{in}),\log(R_\mathrm{char}))$ at the start of the simulations. This is obviously a strong simplification. However, in this paper we are primarily interested in the effects of the new disc evolution paradigm on migration and planetary growth and the formation of embryos \citep[e.g.][]{Voelkel2022,LorekJohansen2022} is beyond the scope of this work. Important parameters are given in Table \ref{tab:pop_init_conds}.

%TTTTTTTTTTTTTT
\begin{table}[!ht]
    \begin{center}
    \caption{Summary of the parameter choice and initial distributions of the variables for the SEPPS.}
    \label{tab:pop_init_conds}
    \begin{tabular}{l c l}
    \hline \hline
    Parameter & Symbol & Value \\ \hline
    Stellar mass & $M_\star$ & $1\,\mathrm{M_\odot}$ \\
    Turbulent viscosity & $\overline{\alpha_{r\phi}}$ & $10^{-4}$ \\
    Active layer & $\Sigma_\mathrm{active}$ & $[1,\,10^{-1},\,10^{-2}]\,\mathrm{g/cm^2}$ \\
    & $\Sigma_\mathrm{active}(f_\mathrm{s})$ & $f_\mathrm{s} \in [0.12,0.08,0.03]$ \\
    Embryo mass & $M_\mathrm{emb}$ & $10^{-2}\,\mathrm{M_\oplus}$ \\
    Ambient FUV field & $\mathcal{F}_\mathrm{FUV}$ & $10\,\mathrm{G_0}$ \\
    \hline
    Variable initial conditions & Symbol \\ \hline
    Disc mass & $M_\mathrm{disc}$ & \\
    Dust to gas ratio & $f_\mathrm{D/G}$ & \\
    Inner radius & $R_\mathrm{in}$ & \\
    Characteristic radius & $R_\mathrm{char}$ & \\
    Wind stress & $\overline{\alpha_\mathrm{\phi z}}$ & \\
    \hline
    \end{tabular}
    \end{center}
\end{table}
%TTTTTTTTTTTTTT

%---------------
\subsection{SEPPS - results} \label{subsec:SEPPS_results}
%---------------

%FFFFFFFFFFFFFF
\begin{figure*}
    \centering
    \includegraphics[width=\textwidth]{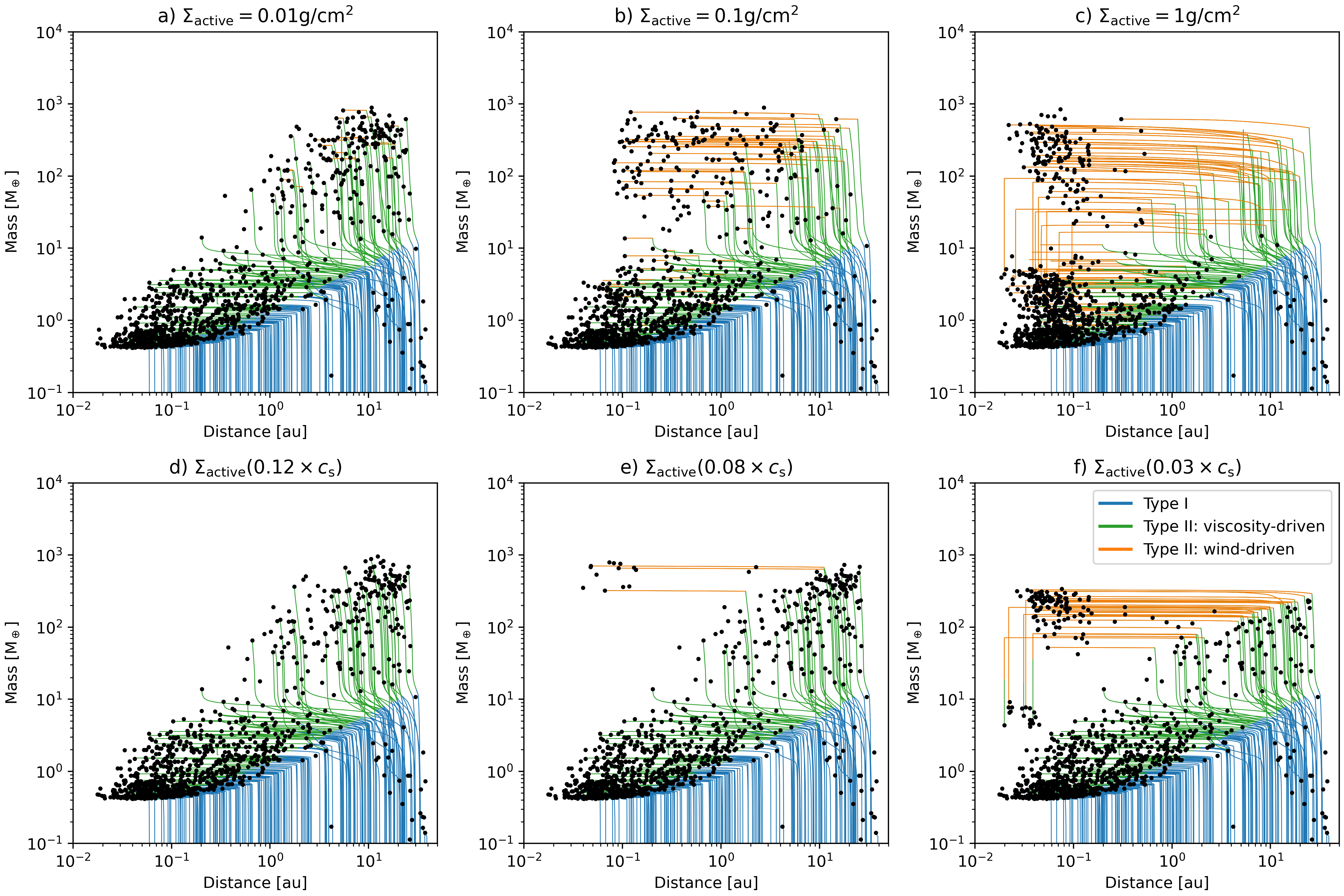}
    \caption{Distance-mass diagrams for single-embryo populations with varying parameterisations of $\Sigma_\mathrm{active}$. The top row shows results for temporally and spatially constant $\Sigma_\mathrm{active}$ while the bottom row shows results with the accretion velocity being a fraction of the local sound speed $f_\mathrm{s}\times c_\mathrm{s}$. Accretion layer thicknesses are increasing from left to right. Formation tracks are shown for 200 randomly selected simulations coloured by migration regime.}
    \label{fig:SEPPS}
\end{figure*}
%FFFFFFFFFFFFFF

We ran populations of $1\,000$ simulations each with varying initial disc conditions and embryo starting locations distributed randomly in log for different parameterisations of $\Sigma_\mathrm{active}$. Simulations were ran until the dispersion of the disc. We would like to point out that the initial conditions of the disc result in disc lifetimes that are in agreement with observations (see panel f) in Fig.~\ref{fig:init_cond}).

Figure~\ref{fig:SEPPS} shows the resulting distance-mass diagram of the different populations along 100 simulation tracks coloured by migration regimes for visualisation. At a very basic level the distance-mass diagrams do not look fundamentally different from observations, i.e. there are many close-in low-mass planets, distant giants and potentially some close-in giants, depending on the accretion layer thickness $\Sigma_\mathrm{active}$.

\begin{figure}
    \centering
    \includegraphics[width=0.48\textwidth]{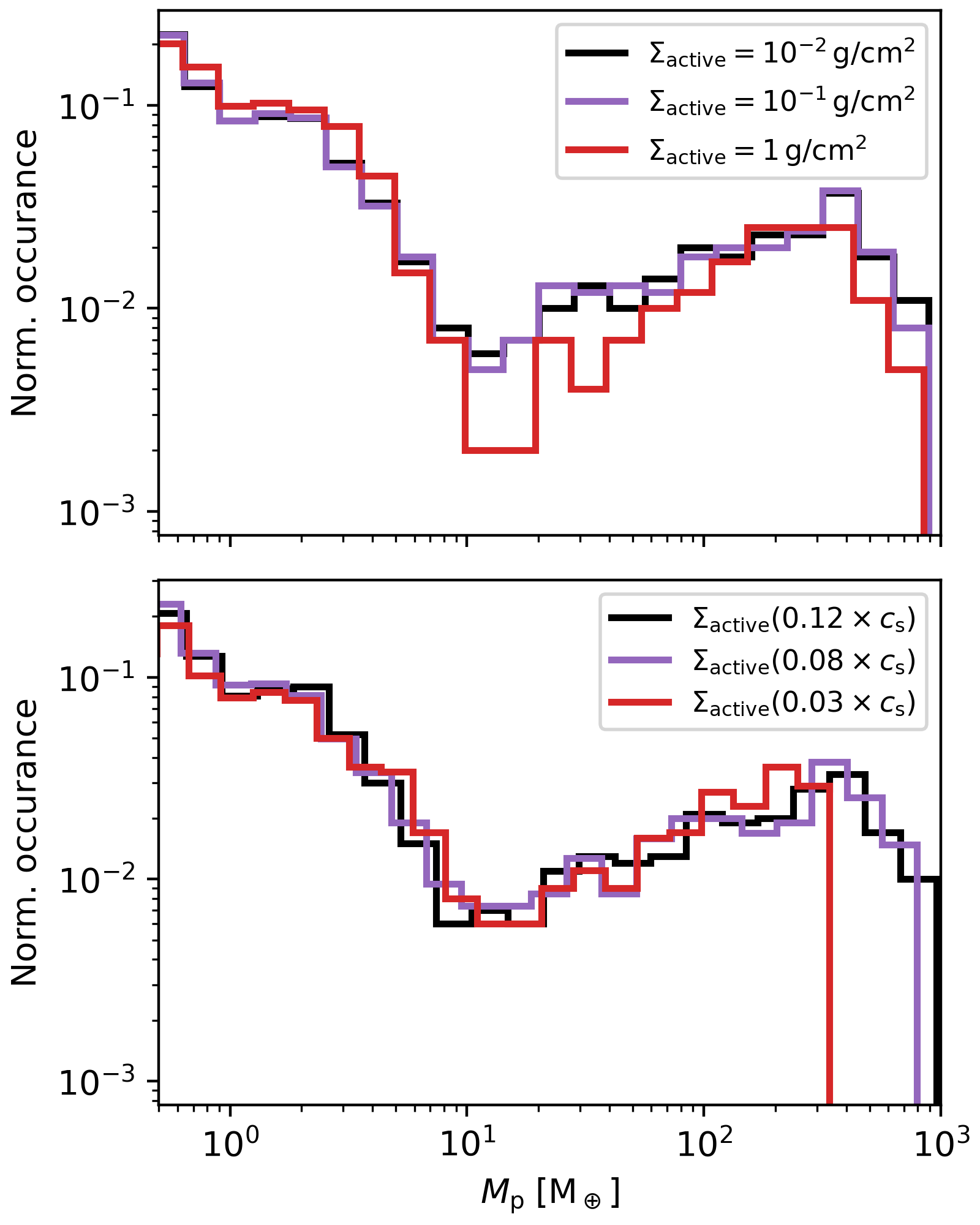}
    \caption{Normalised distribution of planetary masses for the populations shown in Fig.~\ref{fig:SEPPS}. While the overall shape is similar, there are less massive giant planets for slow accretion velocities ($\lesssim 0.08 \times c_\mathrm{s}$).}
    \label{fig:SEPPS_mass_occurrance_rate}
\end{figure}

Based on Fig.~\ref{fig:SEPPS} we make the following observations.
\begin{itemize}
    \item[(i)] For thin accretion layers (i.e. $\Sigma_\mathrm{active}=10^{-2}\,\mathrm{g/cm^2}$ and $f_\mathrm{s}=0.12$), planetary growth halts before they reach the wind-driven Type 2 migration regime. This shows that the planetary mass function at the high end is set by deep gap opening, which makes it difficult to grow past $\sim10^{3}\,\mathrm{M_\oplus}$. This corresponds to the case where the wind-driven regime (Eq.~\ref{eq:wind_driven_disc_dominated}) is ignored and migration is solely due to the viscosity dominated regime (Eq.~\ref{eq:wind_dirven_planet_dominated}).
    
    \item[(ii)] For intermediate layer thickness (i.e. $\Sigma_\mathrm{active}=10^{-1}\,\mathrm{g/cm^2}$ and $f_\mathrm{s}=0.08$) the influence on the giant planet populations starts to emerge with some giant planets migrating far in, before the disc eventually disperses. Note that for $\Sigma_\mathrm{active}(0.08\times c_\mathrm{s})$, $\Lambda_\mathrm{\phi z}$ is remaining at a constant value, and planets eventually enter wind-driven migration at high masses. For the case with $\Sigma_\mathrm{active}=0.1\,\mathrm{g/cm^{2}}$, the critical mass where $|\Lambda_\mathrm{p}|>|\Lambda_\mathrm{\phi z}|$ decreases with $\propto \dot{M}_\mathrm{r,\phi z}$ and hence even planets at masses $\sim 10^{2}\,\mathrm{M_\oplus}$ are influenced.
    
    \item[(iii)] For thick accretion layers (i.e. $\Sigma_\mathrm{active}=1\,\mathrm{g/cm^2}$), the influence is clearly visible. Some of the giant planets intercept the accretion layer during their growth, stop accreting and start migrating inwards rapidly, which can be seen in the planetary mass distribution Fig.~\ref{fig:SEPPS_mass_occurrance_rate}. This leads to many close in giant planets and few to none giant planets outside of $1\,\mathrm{au}$.

    \item[(iv)] For $\Sigma_\mathrm{active}(0.03\times c_\mathrm{s})$ the critical mass is again lower as for panel e) and planets enter the wind-driven migration regime during the runaway growth phase, leading to an upper limit of the mass distribution at about 330 ${M_\oplus}$ in Fig.~\ref{fig:SEPPS_mass_occurrance_rate}.
    
\end{itemize}
Some of the planets with masses $20\,\mathrm{M_\oplus} \lesssim M_\mathrm{p} \lesssim 200\,\mathrm{M_\oplus}$ show a sudden drop in mass due to the loss of their envelope. Young hot giant planets migrated far in are suddenly exposed to direct stellar irradiation $F_\mathrm{irr}$ leading to bloating, driving atmospheric escape $\dot{M} \propto F_\mathrm{irr}/\rho$ in a runaway fashion \citep{Baraffe2004,Sarkis2021,Thorngren2023}, which then leads to the near vertical drop in planetary mass observed in some cases.

With the population syntheses, it is possible to see how physical processes and model parameters like the layer thickness imprint into the demographics of the exoplanet population, which in part can be observed. By running single embryo population, these imprints are visible more clearly. For a detailed comparison with observations, a multi-embryo approach would be needed \citep{BurnMordasini2024}. It is, however, still interesting to qualitatively compare some imprints found here with observations.  

Planetary population syntheses models have predicted a planetary desert \citep{IdaLin2004a} at intermediate masses of $20-200\,\mathrm{M_\oplus}$ that is caused by the high gas accretion rate in the runaway phase. Clear observational evidence of the existence of such a planetary desert is being discussed with some studies pointing towards such a feature \citep{Mayor2011,Bertaux2022,Zang_Microlensing_2025} and others discarding it \citep{Suzuki2018,Bennett2021}. Our simulations show that such a feature persists (see Fig.~\ref{fig:SEPPS_mass_occurrance_rate}). In the simulations presented here, a minimum in the planetary mass function appears around $\sim10-20\,\mathrm{M_\oplus}$ which is lower than what is observed $\sim 40\,\mathrm{M_\oplus}$ \citep{Emsenhuber_NGPPSVII_2025} and what was found in previous planetary population syntheses using viscous disc evolution \citep[e.g.][ NGPPS]{Emsenhuber_NGPPSVII_2025}. In our simulation, the desert is a result of gas runaway accretion, but the location $\sim10-20\,\mathrm{M_\oplus}$ is set by the low pebble isolation and absence of delay of runaway accretion due to higher core masses by planetesimal accretion.

The maximum of the distribution of giant planet masses is at $\sim300\,\mathrm{M_\oplus}$, which is in good agreement with observations, whereas past NGPPS simulations with viscous disks overestimated the giant planet masses \cite{Emsenhuber_NGPPSVII_2025}. On the other hand, we now obtain a sharp drop of the mass function at $\sim 10^{3}\,\mathrm{M_\oplus}$ as already mentioned above, which contrasts observations. In the NGPPS simulations with viscous disks, this upper end was in contrast well reproduced. One can speculate whether additional mechanisms, that are not included here, could allow giant planet growth to higher masses also in low-viscosity disks. A possibility could be the eccentric instability \cite{Papaloizou2001,Kley2006} which could be more easily excited in low-viscosity disks with deep and wide gaps.

Fig.~\ref{fig:SEPPS_mass_occurrance_rate} also shows that although the distance-mass diagrams of panels a) and b) look vastly different, the mass distributions are similar, suggesting that the planets have already stopped growing when they entered the wind-driven Type II migration regime which is possible due to the $\Lambda_\mathrm{\phi z}\propto \dot{M}_\mathrm{r,\phi z}$ dependence of the wind torque. In the case of $\Sigma_\mathrm{active}(f_\mathrm{s}c_\mathrm{s})$ we see that the upper limit of the planetary masses in the population is decreasing with lower $f_\mathrm{s}$, meaning that the planets are entering the wind-driven migration regime during their growth phase.

Theoretical constraints on accretion layer thickness remain sparse. \cite{Mori2021} found values of $\Sigma_\mathrm{active} \approx 0.1-0.6\,\mathrm{g/cm^2}$  and simulations from \citet{Mori_radiative_2025} show slightly subsonic accretion in the active layer. These values are well within the range of where wind-driven Type II migration could play a role for the orbital distance of giant planets, making it an additional mechanism to high eccentricity migration \citep{Wu2003} that can explain close in giant planets.

Interestingly enough, our simulation allow to put independent, observationally motivated constraints on the layer thickness via the comparison with the observed demographics of extrasolar giant planets. It is clear that this constraint must be taken with care due to the various limitations of the model. Still, when taking the six distance-mass diagrams shown in Fig.~\ref{fig:SEPPS} at face value, it is evident that the simulations with the thinnest layer thicknesses (left panels of Fig.~\ref{fig:SEPPS}) are not in agreement with observations as they do not produce any Hot Jupiters. Some of the observed Hot Jupiters that are found close-in very early on are coplanar with the star and have smaller bodies in mean motion resonance which is a clear indication of disk migration leading to their current position \citep{Hord2022}.

Similarly, the thickest layers (right panels of Fig.~\ref{fig:SEPPS}) are in disagreement with observations as they lead to far too many Hot Jupiters compared to the Cold Jupiters (the observed frequency of the former is about $0.5\,\%$, of the latter about $20\,\%$, \citep{Fulton2021}.

The intermediate cases (middle panels of Fig.~\ref{fig:SEPPS}) finally demonstrate that for a distribution active layers with thicknesses that are on the order of $0.1\,\mathrm{g/cm^2}$ (agreeing with \citealt{Mori2021}) or with velocities on the order of $10\,\%$ of the sound speed or possibly a diversity in accretion layer thicknesses, it is possible to obtain a ratio of Hot to Cold Jupiters that is in agreement with observations.

%FFFFFFFFFFFFFF
%FFFFFFFFFFFFFF

%===============
\section{Summary and conclusions} \label{sec:summary_conclusions}
%===============
We have assembled several key aspects of MHD wind-driven discs, including the effects of laminar surface layer accretion on planetary migration and accretion, proposed by \cite{Lega2022} and \cite{Nelson2023}, into a global planet formation model and performed single-embryo planetary population syntheses within this new disc evolution paradigm. We investigated the impact of different parameterisations for the thickness of the accreting layer $\Sigma_\mathrm{active}$.

We present the following conclusions based on our model:
\begin{enumerate}
    \item On a very basic level, the paradigm of layered MHD-wind driven discs with planet formation via core accretion including pebble accretion leads to populations that share the key properties of the observed planet population (numerous close-in low-mass and outer, less numerous giant planets, the location of which depends on layer thickness.) 

    \item The thickness of the active layer has a large impact on the final location of giant planets. The thicker the layer, the more giant planets migrate close to their host star. Fast, wind-driven Type II migration could pose an alternative origin to close in giant planets in addition to high eccentricity migration. 

    \item A detailed quantitive comparison with the exoplanet population is beyond the scope of this work and would require to include more physical processes. But if we take the results obtained here at face value, then one finds that the resulting synthetic distance-mass diagrams are in agreement with the observed one in terms of the relative frequency of Hot versus Cold Jupiters for a layer thickness on the order of $0.1\,\mathrm{g/cm^2}$ or velocity of approximately $10\,\%$ of $c_{\rm s}$. Clearly thinner or faster layers are incompatible because no Hot Jupiters form at all and thicker/slower layers lead on the opposite to too many close-in giants.   

    \item Time and location of the critical mass for entering wind-driven Type II migration poses a hard limit to upper end of the final mass distribution of giant planets, favouring also rather fast $\gtrsim 0.1\times c_\mathrm{s}$ and thin accretion layers $\gtrsim0.1\,\mathrm{g/cm^2}$.

    \item In the absence of wind-driven Type II migration (i.e. for very thin accretion layers), we find Type II migration to be inefficient due to deep gap opening, leading to a bifurcation of low mass planets forming further in and high mass planets forming further out in the disc. This means in particular that in low-viscosity disks without an active layer at all (or a very thin one), Hot Jupiter formation via disk migration is not possible. Giant planets grow, once they have started runaway gas accretion, virtually in situ. 

    \item Using current gap opening criteria and gas accretion rates derived from hydrodynamic simulations, it is difficult to grow past $\gtrsim\,10^{3}\,\mathrm{M_\oplus}$ in low viscosity ($\overline{\alpha_\mathrm{r\phi}}=10^{-4}$) due to efficient and deep gap opening. This is in contrast to the observed upper end of the planetary mass function. Alternative physical mechanisms not included here, like potentially the eccentric instability \citep{Papaloizou2001,Kley2006} may allow growth to higher masses.
\end{enumerate}
Many questions remain regarding planet formation in MHD wind-driven discs that have to be addressed using 2D and 3D magnetohydrodynamic simulations and that should then be broken down to 1D for inclusion in global models allowing statistical comparison with observations. A key question is the formation and depth of gaps and how planets located in these gaps accrete gas. Furthermore, Type II migration is closely related to the gap formation process and remains highly uncertain with results differing between models \citep{Lega2022,AoyamaBai2023,Wafflard-Fernandez2023}. Finding analytic expressions for the migration rates (e.g. outward migration found by \citealt{Wafflard-Fernandez2023}) would allow us to test the implications of such migration behaviour on planet formation on a global view, similarly as done in this work.

Our simulations already show some distinctions from synthetic populations in viscous discs, and draw a picture similar to what was proposed in \cite{Ziampras2025}.

In this work we neglected any effects from the radial dependence and time evolution of the wind torque $\overline{\alpha_\mathrm{\phi z}}$ and turbulent viscosity $\overline{\alpha_\mathrm{r\phi}}$ that can lead to inner cavities and migration traps \citep[e.g. ][]{Ogihara2015b,Flock2019,Speedie2022,Alessi2022}. These strongly depend on the thermal physics of the disc and the magnetic field configuration where the latter remains poorly constrained. Testing existing constraints \citep[e.g.][]{Lesur2021,Kobayashi2025} using planetary population syntheses approach will give insights that will be valuable to assess our current knowledge. This will be subject to future studies.

% ----------------
% acknowledgements
% ----------------
\begin{acknowledgements}
We would like to thank Elena Lega, Aurélien Crida, Alessandro Morbidelli and Richard Nelson for fruitful discussions. We would also like to thank Shoji Mori for providing more insights on the results on accretion layer from 2D magnetohydrodynamic simulations. We would like to thank the referee for their useful comments that greatly helped improving the clarity of the manuscript. We further thank Alessandro Morbidelli for a thorough reading of a first version of this manuscript. J.W. and C.M. acknowledge the support from the Swiss National Science Foundation under grant 200021\_204847 “PlanetsInTime”. Part of this work has been carried out within the framework of the NCCR PlanetS supported by the Swiss National Science Foundation under grants 51NF40\_182901 and 51NF40\_205606. Calculations were performed on the Horus cluster of the Division of Space Research and Planetary Sciences at the University of Bern.
\end{acknowledgements}

\bibliographystyle{bibtex/aa.bst}
\bibliography{references.bib}

\onecolumn

\begin{appendix}
    % ==================
    \section{Distributions of the initial conditions for the population and resulting disc lifetime distribution} \label{app:init_dist}
    % ==================
    
    %FFFFFFFFFFFFFFF
    \begin{figure*}[h!]
        \centering
        \includegraphics[width=\hsize]{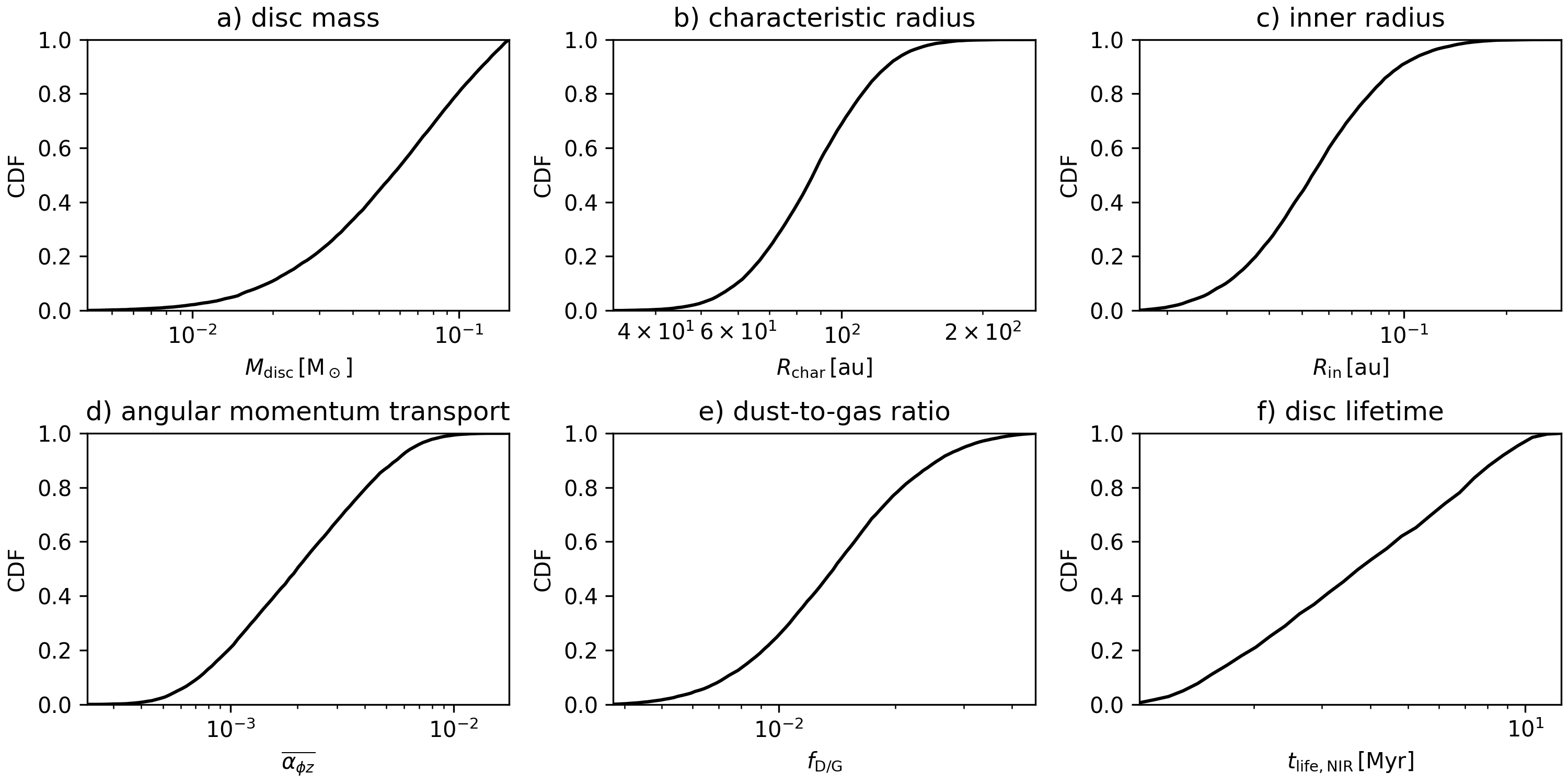}
        \caption{Distributions of the initial disc conditions according to Sect.~\ref{subsec:SEPPS_init_conds}. Resulting NIR disc lifetimes according to \cite{Kimura2016} are shown in panel f).}
        \label{fig:init_cond}
    \end{figure*}
    %FFFFFFFFFFFFFFF

\end{appendix}

\end{document}